\documentclass{jpsj-suppl}
\usepackage{txfonts} 

\title{Nuclear Physics and Astrophysics of Neutrino Oscillations}

\author{A.B. \textsc{Balantekin}}

\inst{Department of Physics, University of Wisconsin, Madison, WI 53706, USA}

\email{baha@physics.wisc.edu}

\recdate{September 2, 2016}

\abst{For a long time very little experimental information was available about neutrino properties, even though a minute neutrino mass has intriguing cosmological and astrophysical implications. This situation has changed in recent decades: intense experimental activity to measure many neutrino properties took place. Some of these developments and their implications for astrophysics and cosmology are briefly reviewed with a particular emphasis on neutrino magnetic moments and collective neutrino oscillations.}

\kword{Collective neutrino oscillations, neutrino magnetic moments, r-process nucleosynthesis}

\begin{document}
\maketitle

\section{Introduction}

Nuclear astrophysics aims to understand the heavens using nuclear physics as a tool. Progressively more capable satellites COBE, WMAP, Planck and powerful telescopes such as Subaru and Keck, with even more powerful telescopes coming on line such as the Thirty Meter Telescope made phenomena where the microphysics is nuclear and particle physics increasingly accessible. Aside from ``practical" consequences such as unraveling the origin of the elements around us, these efforts do lead to a deeper appreciation of the night sky in a centuries old tradition:  In 1689 the haiku poet Basho noted the majesty of the Milky Way enveloping the Sado island off the Echigo coast (present day Niigata), where this meeting (2106 Nuclei in Cosmos conference) is auspiciously located. 
 
In many cases the bridge between observational astronomy and laboratory experiments is provided by neutrinos. These particles indeed play a very special role in the Cosmos: since they interact only weakly they can transport energy and entropy over large distances. In addition, since they can flip the third component of the strong isospin, they control the electron fraction in environments where nucleosynthesis takes place. 

More than half a century after their existence was first postulated, we finally seem to be getting closer to understanding the elusive physics of neutrinos. For a long time very little experimental information was available about neutrino properties, even though a minute neutrino mass has intriguing cosmological and astrophysical implications. This situation has changed in recent decades: intense experimental activity to measure many neutrino properties took place. For example it is now well established that weak-interaction eigenstates of the neutrinos do not coincide with the mass eigenstates, but are related by a unitary transformation, which can be parameterized as 
\begin{equation}
\label{1a}  
\left(
\begin{array}{ccc}
 1 & 0  & 0  \\
  0 & C_{23}   & S_{23}  \\
 0 & -S_{23}  & C_{23}  
\end{array}
\right)
\left(
\begin{array}{ccc}
 C_{13} & 0  & S_{13} e^{-i\delta_{CP}}  \\
 0 & 1  & 0  \\
 - S_{13} e^{i \delta_{CP}} & 0  & C_{13}  
\end{array}
\right) 
\left(
\begin{array}{ccc}
 C_{12} & S_{12}  & 0  \\
 - S_{12} & C_{12}  & 0  \\
0  & 0  & 1  
\end{array}
\right)
\end{equation}
where $C_{ij} = \cos \theta_{ij}$, $S_{ij} = \sin \theta_{ij}$, and $\delta_{CP}$ is the CP-violating phase. Already a few years back the angles $\theta_{12}$ and $\theta_{23}$ were determined from the solar, atmospheric, and accelerator experiments, a fact that was recognized with the 2015 Nobel Prize in physics. Within the last decade, three reactor neutrino experiments, Daya Bay \cite{An:2012eh}, RENO \cite{Ahn:2012nd}, and Double Chooz \cite{Abe:2011fz}, were also able to measure the remaining mixing angle $\theta_{13}$. At short baselines the electron antineutrino
survival probability is given as
\begin{equation}
P(\overline{\nu}_e \rightarrow \overline{\nu}_e) = 1 - \sin^2 2 \theta_{13} \sin^2 \left( \frac{\delta m_{ee}^2L}{4E} \right) 
- \cos^4 \theta_{13} \sin^2 2\theta_{12} \sin^2 \left( \frac{\delta m_{21}^2 L}{4E} \right)
\end{equation}
where the quantity $\delta m_{ee}^2$ is defined via the equality
\begin{eqnarray}
\sin^2 \left( \frac{\delta m_{ee}^2L}{4E} \right) &=& \cos^2 \theta_{12} \sin^2 \left( \frac{\delta m_{31}^2L}{4E} \right) + 
\sin^2 \theta_{12} \sin^2 \left( \frac{\delta m_{32}^2L}{4E} \right) \nonumber \\
&=& \sin^2 \left( \frac{\delta m_{31}^2L}{4E} \right) - \sin^2 \theta_{12} \sin^2 \left( \frac{\delta m_{21}^2L}{4E} \right) .
\end{eqnarray}
Note that at short distances the second term in the second line is extremely small. A recent measurement of the inverse beta decay with the full detector configuration at Daya Bay results in \cite{An:2015rpe} $|\delta m_{ee}^2| = (2.42 \pm 0.11) \times 10^{-3}$ eV$^2$ and $\sin^2 2 \theta_{13} = 0.084 \pm 0.005$. A simultaneous fit to the neutron capture on hydrogen and neutron capture on Gd at Daya Bay yields $\sin^2 2 \theta_{13} = 0.082 \pm 0.004$. \cite{An:2016bvr}. An independent measurement at RENO \cite{Kim:2016yvm} gives $|\delta m_{ee}^2| = [2.62 ^{+0.21}_{-0.23} (\rm stat.) 
^{+0.12}_{-0.13} (\rm syst.) ] \times 10^{-3}$ eV$^2$ and $\sin^2 2 \theta_{13} = 0.082 \pm 0.009 (\rm stat.)\pm 0.006 (\rm syst.) $.  
These results are also in agreement with the measurements of the Double Chooz collaboration \cite{Abe:2015rcp}.
Thus not only different experiments measuring $\theta_{13}$ converge on the same value, but they also single out this angle as the most precisely known neutrino mixing angle. 

\section{Reactor anomaly}

Even though recent reactor neutrino experiment discussed above were very successful in completing the measurement of the neutrino mixing matrix, they introduced a new puzzle. Predicting the reactor antineutrino flux is a very difficult task \cite{Hayes:2016qnu,Huber:2011wv}, 
very few of the matrix elements describing beta decays of the isotopes in the reactor fuel mix can be calculated to the desired accuracy. 
But already a few years back it was remarked that the measured reactor neutrino flux is about 5\% less that the predicted one \cite{Mention:2011rk}. In addition 
to this overall deficiency precise measurements of the reactor neutrino flux show an increase in the flux around a neutrino energy of $\sim 5$ MeV \cite{An:2015nua,Kim:2016yvm}. This ``bump" and some other anomalies observed at neutrino experiments can be interpreted as one or more sterile neutrinos mixing with active ones \cite{Abazajian:2012ys}.
Many interesting nuclear physics aspects of the problem, such as the contributions from forbidden transitions, notwithstanding the size and the shape of the reactor antineutrino spectra could be best addressed with dedicated experiments  \cite{Ashenfelter:2015uxt}. Recent IceCube \cite{TheIceCube:2016oqi} and Daya Bay \cite{An:2014bik,An:2016luf}
results significantly shrink the parameter space for sterile states, but they cannot completely rule it out. 
In fact, a recent joint analysis of the disappearance searches in the MINOS, Daya Bay, and Bugey-3 experiments \cite{Adamson:2016jku}
still allows part of the sterile neutrino parameter space hinted by the LSND and MiniBooNE experiments \cite{Conrad:2013mka}. Whether one can interpret the reactor anomaly as active-sterile neutrino mixing is still an open question \cite{Giunti:2015wnd}. However, as we discuss below, a light sterile neutrino would impact neutrino magnetic moment measurements at reactors. 

\section{The impact of neutrino magnetic moment in astrophysics, cosmology, and nucleosynthesis}

At lower energies, beyond Standard Model physics is described by local operators resulting in the effective Lagrangian 
\begin{equation}
{\cal L}_{\rm eff} = {\cal L}_{\rm SM} + \frac{C^{(5)}}{\Lambda} {\cal O}^{(5)} + \sum_i \frac{C_i^{(6)}}{\Lambda^2}
{\cal O}_i^{(6)} + \sum_i \frac{C_i^{(7)}}{\Lambda^3} 
{\cal O}_i^{(7)}  + \cdots
\label{a1}
\end{equation}
The second term in the right-hand side of the Eq. (\ref{a1}) is the unique dimension five operator which can be written using the Standard Model fields and it represents the Majorana mass term in the Lagrangian. The fourth term includes the neutrino magnetic moment. The best experimental limits on neutrino magnetic moment come from reactor experiments measuring the recoil energy of the electrons struck by 
antineutrinos. The cross section for this reaction is given by  
\begin{equation}
\label{cs}
\frac{d \sigma}{dT_e} = \frac{\alpha^2 \pi}{m_e^2} \mu_{\rm eff}^2 \left[ \frac{1}{T_e} - \frac{1}{E_{\nu}} \right]
\end{equation}
where the effective magnetic moment 
\begin{equation}
\label{mueff}
\mu_{\rm eff}^2 = \sum_i \left| \sum_j U_{ej} e^{-i E_jL} \mu_{ji} \right|^2
\end{equation}
takes into account the oscillation of the neutrinos as they travel a distance $L$ from the reactor core to the detector. 
In Eq. (\ref{cs}), $E_{\nu}$ is the energy of the incoming neutrino and $T_e$ is the kinetic energy of the recoil electron. 
Since this cross section exceeds the standard weak-interaction cross section at low recoil energies, the lowest recoil energy accessible in an experiment determines the upper limit on the magnetic moment. Currently the best experimental upper limit is  $2.9 \times 10^{-11} \>  \mu_B$ at 90\% C.L. \cite{Beda:2013mta}. 
Astrophysical and cosmological considerations also put limits on the neutrino magnetic moment. For example, a large enough neutrino magnetic moment implies an enhanced rate for the decay of stellar plasmons into neutrino-antineutrino pairs. Since the neutrinos freely escape the star, this is turn cools a red giant star faster, delaying helium ignition. The current best such limit comes from globular cluster M5. It is given as 
$ 4.5 \times 10^{-12} \> \mu_B$ \cite{Viaux:2013hca}. It is also possible to place limits on neutrino magnetic moments from cosmological considerations. For example, requiring synthesis of $^4$He in the Big Bang Nucleosynthesis (BBN) epoch in the Early Universe not be disrupted by the excitation of additional neutrino helicity states limits the value of neutrino magnetic moment \cite{Morgan:1981zy}. This 
last argument, however, applies only to Dirac neutrinos since for Majorana neutrinos no new states are created in the interactions via the magnetic moment. 

Because of the factor $L$ in Eq. (\ref{mueff}), the effective neutrino magnetic moment measured at close distances to reactors will be different that that is measured using solar neutrinos \cite{Balantekin:2014mqa}. In reactor experiments, if the 5 MeV enhancement is indeed due to a sterile state oscillating into an active one, the relevant phase in Eq. (\ref{mueff}) will average to zero and we obtain  \cite{Balantekin:2013sda}
\begin{equation}
\mu_{\rm eff}^2 \le \sum_{i=1}^3 \mu_{i4}^2 + \left( 1 - |U_{e4}|^2 \right) \sum_{i,j=1}^2 \mu_{ij}^2 \> .
\end{equation}
Consequences of this effect are unfortunately not likely to be observable in the near future. 

Neutrino magnetic moments are not zero in the Standard Model, but they are very small: order of $\sim 10^{-20} \> \mu_B$ \cite{Balantekin:2013sda}. However, beyond the Standard Model physics may bring them closer to the experimental and observational limits of $10^{-11} - 10^{-12} \> \mu_B$. Should this be the case the weak decoupling at the BBN epoch should be reconsidered since the magnetic cross sections have a rather different energy dependence than the weak cross sections. It turns out that light element abundances and other cosmological parameters are sensitive to magnetic couplings of the order of $10^{-10} \>\mu_B$ if the neutrinos are Majorana type \cite{Vassh:2015yza}. 
(For Dirac neutrinos magnetic interactions produce right-handed states which decouple from the thermal equilibrium). One example is the impact of the Majorana magnetic moment on the relativistic energy density in thermal equilibrium, which is usually employed to define the quantity $N_{\rm eff}$: 
\begin{equation}
\rho_{\rm relativistic} = \frac{\pi^2}{15} T_{\gamma}^4 \left[ 1 + \frac{7}{8} N_{\rm eff} \left(\frac{4}{11}\right)^{4/3} \right] .
\end{equation}
Using the value $N_{\rm eff}=3.30 \pm 0.27$ quoted by the Planck collaboration \cite{Ade:2013zuv}, one can then place a limit of $\mu_{\rm Majorana} \le 6 \times 10^{-10} \mu_B$ for the Majorana magnetic moment \cite{Vassh:2015yza}. This complements the limits on the Dirac neutrino magnetic moment derived earlier using BBN considerations. 

\section{Neutrino Collective Oscillations} 

Neutrinos play a key role in many astrophysical and cosmological phenomena \cite{Balantekin:2013gqa,Bertulani:2016eru}. In a core-collapse supernova 99\% of the emitted energy is carried away by neutrinos. The dynamics of such supernovae is essentially controlled by neutrinos.  As discussed below even though the site of r-process nucleosynthesis is still an open question, in all possible sites neutrinos control the value of the electron fraction, the parameter determining the yields of the r-process. 

The site of r-process nucleosynthesis is not yet identified. Two leading candidates, core-collapse supernovae and neutron star mergers, both have features that need to be better understood.  Observation of elemental abundances in metal-poor stars help us to probe the nucleosynthetic processes that created those elements \cite{Jacobson:2013tba}. Observational astronomy is making excellent progress: 
chemical signatures of first-generation very massive stars are indeed observed recently for the first time \cite{aoki}. For neutron stars to merge they need to be first formed following the evolution of earliest stars.  Furthermore if the rates of occurrence of neutron star mergers is low it could lead to r-process enrichment that is not consistent with observations at these very low metallicities \cite{Argast:2003he}. It should be noted that recently several orders of magnitude greater enhancement of r-process element abundances was observed in an ultra-faint dwarf (i.e. very old) galaxy than that seen in other such galaxies, implying that a single rare event produced the r-process material \cite{ji}, an argument in favor of neutron star mergers. Another signature of the neutron-star mergers may be looking for the electromagnetic transients from the decay of radioactive isotopes they would produce  \cite{Martin:2015hxa}. The LIGO experiment has at least twice detected gravitational waves from the merger of two black holes. Neutron star mergers also produce gravitational waves. Their future detection will  help evaluate neutron star merger rate and hence the distribution of elements in the Cosmos. It should also be stressed that a complete understanding of the r-process nucleosynthesis would require crucial input from nuclear physics \cite{Balantekin:2014opa} as individual nuclear properties such as nuclear masses as well as decay and capture rates still have significant uncertainties \cite{Mumpower:2015ova,Martin:2015xql}. 

The large number of neutrinos cooling the proto-neutron star in a core-collapse supernova necessitates including the effects of neutrino-neutrino interactions in the description of neutrino transport. To illustrate this effect let us assume that there are only two- neutrino flavors and introduce 
the generators of the neutrino flavor isospin algebras  \cite{Balantekin:2006tg}: 
\begin{eqnarray}
J^+_{{\bf p}} &=& a_x^\dagger({\bf p}) a_e({\bf p}), \> \> \>
J^{\>-}_{{\bf p}}=a_e^\dagger({\bf p}) a_x({\bf p}), \nonumber \\
J^0_{{\bf p}} &=& \frac{1}{2}\left(a_x^\dagger({\bf p})a_x({\bf p})-a_e^\dagger({\bf p})a_e({\bf p}) 
\right). \label{su2}
\end{eqnarray}
where appropriate neutrino creation and annihilation operators are introduced.  Defining the auxiliary vector quantity in terms of the neutrino mixing angle 
\begin{equation}
\mathbf{B} = (\sin2\theta,0,-\cos2\theta), 
\end{equation}
the total Hamiltonian with two flavors,containing one- and two-body interaction terms, can be written as
\begin{equation}
\label{total}
\hat{H}_{\mbox{\tiny total}} = H_{\nu} + H_{\nu \nu} 
= \left(
\sum_p\frac{\delta m^2}{2p}\mathbf{B}\cdot\mathbf{J}_p  - \sqrt{2} G_F 
N_e  J_p^0  \right) 
+ \frac{\sqrt{2}G_{F}}{V}\sum_{\mathbf{p} \neq\mathbf{q}}\left(1- 
\cos\vartheta_{\mathbf{p}\mathbf{q}}\right)\mathbf{J}_{\mathbf{p}}\cdot\mathbf{J}_{\mathbf{q}}  
\end{equation} 
where $\cos\vartheta_{\mathbf{p}\mathbf{q}}$ is the angle between neutrino momenta $\mathbf{p}$ and $\mathbf{q}$. Usually the term containing this angle is averaged over in an approximation known as the {\em single angle approximation}:
\begin{equation}
\label{satotal}
\hat{H}_{\mbox{\tiny SA}}   
= \left(
\sum_p\frac{\delta m^2}{2p}\mathbf{B}\cdot\mathbf{J}_p  - \sqrt{2} G_F 
N_e  J_p^0  \right) 
+ \frac{\sqrt{2}G_{F}}{V}   \left(\langle 1- 
\cos\vartheta_{\mathbf{p}\mathbf{q}} \rangle \right)  \sum_{\mathbf{p} \neq \mathbf{q}}     \mathbf{J}_{\mathbf{p}}\cdot\mathbf{J}_{\mathbf{q}}  .
\end{equation} 
The eigenstates of the Hamiltonian in Eq. (\ref{satotal}) 
\begin{equation}
| x_i \rangle = \prod_{i=1}^N \sum_k \frac{2 k \mathbf{J}^{\dagger}_{\mathbf{k}}}{\delta m^2 - 2k x_i} |0 \rangle
\end{equation}
can be obtained \cite{Pehlivan:2011hp} by solving the Bethe ansatz equations for the variables $x_i$ 
\begin{equation}
- \frac{V}{\sqrt{2}G_F \langle 1- \cos\vartheta \rangle} - \sum_p \frac{2p j_p}{\delta m^2 - 2p x_i} = \sum_{j \neq i} \frac{1}{x_i-x_j} .
\end{equation}
In addition one can identify the conserved quantities for each momenta. In mass basis they are 
\begin{equation}
\label{invar}
\mathbf{h}_p = \frac{2\sqrt{2}G_Fp}{V\delta m^2} \langle 1- \cos\vartheta \rangle \sum_{q \neq p} \left( \frac{q}{q-p}  \mathbf{J}_{\mathbf{p}} \cdot 
\mathbf{J}_{\mathbf{q}} \right) + \mathbf{J}^0_{\mathbf p}.
\end{equation}
The eigenvalues of these conserved quantities are also given in terms of the Bethe ansatz solutions, $x_i$. Note that these invariants can vanish for some values of neutrino momenta. The coefficient in front of the sum in Eq. (\ref{invar}) is proportional to the ratio of the vacuum oscillation length scale, $L_v^{-1} \sim \delta m^2/ p$, and the self interaction length scale, $ L_s^{-1} \sim G_F  \langle 1- \cos\vartheta \rangle /V$. After  each invariant is multiplied by $\delta m^2/2p$, the sum  gives the neutrino transport Hamiltonian itself. 

One can incorporate antineutrinos in this formalism  by introducing a second set of SU(2) algebras. Similarly three flavors require introduction of two sets of SU(3) algebras. 

If there are no sterile neutrinos the CP-violating phase, $\delta_{CP}$, can be factored out of the one-body neutrino-mixing Hamiltonian 
\cite{Balantekin:2007es}. This result also holds for the single-angle collective neutrino Hamiltonian in the mean field limit \cite{Gava:2008rp} as well as in the full two-body case \cite{Pehlivan:2014zua} provided that there are no neutrino magnetic moments interacting with an external magnetic field. If there are external magnetic fields, the neutrino magnetic moment matrix should be replaced by an effective one  including the CP violating phase. 

Many times the single-angle Hamiltonian, Eq. (\ref{satotal}), is solved in the mean field approximation \cite{Duan:2010bg}. In this approximation the two-body term is replaced by a one-body term:
\begin{equation}
\mathbf{J}_{\mathbf{p}}\cdot\mathbf{J}_{\mathbf{q}} \rightarrow \langle \mathbf{J}_{\mathbf{p}} \rangle \cdot\mathbf{J}_{\mathbf{q}} 
+ \mathbf{J}_{\mathbf{p}}\cdot \langle  \mathbf{J}_{\mathbf{q}}  \rangle 
\end{equation}
where the averaging is done over an appropriately chosen state. The resulting Hamiltonian then represents a test neutrino interacting with a mean field which describes the effect of all the other neutrinos on that test neutrino. 
For the averaging usually an SU(2) coherent state is chosen \cite{Balantekin:2006tg}. 
Once can additionally include the neutrino-antineutrino pairing field \cite{Serreau:2014cfa} which would give a mean field proportional to the neutrino masses. 
Such a term should play a role in anisotropic environments \cite{Cirigliano:2014aoa}. 
Several physical properties of the many neutrino gas can be explored in the mean-field approximation. For example, the electron fraction resulting from the collective oscillations exhibit an oscillatory behavior in certain cases 
\cite{Balantekin:2004ug}. Another interesting effect resulting from studying the collective neutrino oscillations in the mean-field approximation is spectral swappings or splits, on the final neutrino energy spectra: at a particular energy these spectra are almost completely divided into parts of different flavors \cite{Raffelt:2007cb,Duan:2008za}. The validity of the mean-field approximation was recently tested in Ref. 
\cite{Pehlivan:2016voj}  where the adiabatic solution of the exact single-angle Hamiltonian was presented with two flavors in inverted hierarchy and in a calculation with 250 neutrinos only. A spectral split is also observed. It should also be noted that appropriately defined averages of the invariants in Eq. (\ref{invar}) become constants of motion in the mean-field approximation. 

\section{Conclusions}

 Their seemingly very small masses and feeble interactions with ordinary matter make neutrinos rather special. We made a lot of progress in understanding the elusive physics of neutrinos. Much experimental information is now available with more to come about neutrino properties enabling us to use them as valuable tools in exploring the Cosmos. These tools will be put into a good use with the upcoming powerful observatories and laboratory facilities.

\section*{}
I would like to thank members of the National Astronomical Observatory of Japan for their hospitality during my many visits. 
This work was supported in part by the US National Science 
Foundation Grant No. PHY-1514695 and and 
in part by the University of Wisconsin Research Committee with funds 
granted by the Wisconsin Alumni Research Foundation.

\end{document}